\documentclass{WileyChemistry-template}

\makeatletter
\DeclareOldFontCommand{\rm}{\rmfamily}{\mathrm}
\makeatother

\title{\raggedright Ladder-like Structural Architecture of Layered Magnetic \textit{A}\textsubscript{2.4}\ce{Cr8Te14} (\textit{A} = Rb, Cs) Compounds by Self-flux Synthesis}

\author{
\begin{minipage}{\textwidth}
	Kai D. Röseler,\textsuperscript{[a]} Felix Eder,\textsuperscript{[a]} Fabian O. von Rohr,*\textsuperscript{[a]}
\end{minipage}
}

\newcommand{\affiliation}{
\begin{itemize}

\item[{[a]}] K. D. Röseler, Dr. F. Eder, Prof. Dr. F. O. von Rohr*\\
Department of Quantum Matter Physics, University of Geneva, CH-1211 Geneva, Switzerland%\\
\end{itemize}
}

\renewcommand{\abstract}{The discovery and control of intergrowth structures represent an important avenue for the targeted synthesis of new, more complex structure types. When including magnetic framework metal atoms, this enhanced complexity can transfer to rich magnetic ground states. Here, we show that the subtle adjustment of the composition of alkali-tellurium fluxes enables the synthesis of a new family of alkali chromium tellurides, \textit{A}\textsubscript{2.4}\ce{Cr8Te14} (\textit{A} = Rb, Cs). Their ladder-like crystal structures integrate the two-dimensional character of delafossite-like $A$\ce{CrTe2} with the tunnel motifs of hollandite-like $A_{x}$\ce{Cr5Te8} phases. This results in a previously unobserved unique hybrid framework. Direction-dependent magnetization measurements on oriented single crystals reveal distinct magnetic ground states: \ce{Rb_{2.4}Cr8Te14} is antiferromagnetic with $T_{\rm N}$ = 114.5 K, while \ce{Cs_{2.4}Cr8Te14} is ferrimagnetic with $T_{\rm C}$ = 125.0 K. This work underscores the simplicity and effectiveness of flux growth as a design strategy for discovering low-dimensional materials.}

\newcommand{\keywords}{
	chromium tellurides \textbullet\ 
	intergrowth phases \textbullet\ 
	layered materials \textbullet\ 
	flux synthesis \textbullet\ 
	low-dimensional magnetism
}

\begin{document}\sloppy
%%%%%%%%%%%%%%%%%%%%%%%%%%%%%%%%%%%%%%%%%%%%%%%%%%%%%%%%%%
%%%%%%%%%%%%%%%%%%%%%%%%%%%%%%%%%%%%%%%%%%%%%%%%%%%%%%%%%%
%%%%%%%%%%%%%%%%%%%%%%%%%%%%%%%%%%%%%%%%%%%%%%%%%%%%%%%%%%

\twocolumn[\vspace{-1.5cm}\maketitle\vspace{-1cm}
	\textit{\dedication}\vspace{0.4cm}]
\small{\begin{shaded}
		\noindent\abstract
	\end{shaded}
}

\begin{figure} [!b]
\begin{minipage}[t]{\columnwidth}{\rule{\columnwidth}{1pt}\footnotesize{\textsf{\affiliation}}}\end{minipage}
\end{figure}

%%%%%%%%%%%%%%%%%%%%%%%%%%%%%%%%%%%%%%%%%%%%%%%%%%%%%%%%%%
%%%%%%%%%%%%%%%%%%%%%%%%%%%%%%%%%%%%%%%%%%%%%%%%%%%%%%%%%%
%%%%%%%%%%%%%%%%%%%%%%%%%%%%%%%%%%%%%%%%%%%%%%%%%%%%%%%%%%

%%%%%%%		 Main Text			%%%%%%% 

%	For Communications for Angewandte Chemie, please remove headlines for Introduction, Results and Discussion and Conclusion

\section*{Introduction}
\label{introduction}
%Information on our Journals can be found on the websites of \href{https://chemistry-europe.onlinelibrary.wiley.com/}{\textit{Chemistry Europe}} and

The search for new functional materials increasingly turns to modular “synthesis-by-design” strategies, in which known structural units are deliberately combined to generate targeted properties.\cite{jansen2002concept,chamorro2018progress,cheetham2022chemical} One structural realization of this principle is the formation of complex structures of compounds that can be interpreted as intergrowth of segments of simpler structures.\cite{Parthe_1985_Series,nesper1991bonding} Such intergrowth phases often serve as structural bridges between known phases and provide a rational route toward new, more complex architectures.\cite{Kanatzidis_2005_Homologies,zhao2025stoichiometrically}

For this study, we considered the family of ternary alkali-chromium tellurides. In this system, all known phases up to date follow one of two structure types: The first known structure type is found for the smaller alkali cations \textit{A} = Li, Na, K and can be considered as delafossite-like layered phases with a composition of \textit{A}\ce{CrTe2}. The crystal structure of these delafossite-type phases consists of \ce{CrTe2} sheets separated by layers of intercalated alkali cations.\cite{kobayashi2016_LiCrTe2_NaCrTe2,Huang_2022_NaCrTe2} We have recently established a synthetic approach to obtain large single-crystals of delafossite-type \ce{ACrX2} (X = chalcogen) phases by employing a \textit{A}/X-self-flux synthesis approach.\cite{Witteveen2023,Eder2025,eder2025stoichiometry} It is worth noting that the deintercalation of these phases by soft-chemical methods has become an central strategy for accessing metastable, magnetic, Cr\textsuperscript{IV}-based vdW materials such as \ce{CrTe2}.\cite{Freitas_2015,Roeseler_CrTe2}\\

The second known structure type for \textit{A}--Cr--Te phases -- exhibited by the heavier alkali metals -- follows a composition of $A_{x}$\ce{Cr5Te8} (\textit{A} = K, Rb, Cs; $0.73\,\leq\,x\,\leq\,1$) and corresponds to a hollandite-like architecture. Here, the \ce{CrTe2} layers are interconnected by \ce{Cr2Te2} slabs forming a tri-periodic framework perforated by channels occupied by alkali cations.\cite{Yamazaki2011,Bouteiller2023} 
The synthesis route reported for these phases either used a solid-state reaction of the elements\cite{Yamazaki2011} (\textit{A} = K, Rb, Cs), or a reaction of elemental Cr, Te and either \ce{Cs2CO3} or \ce{CsTe4} as Cs-sources\cite{Bouteiller2023}. By altering the amount of Cs-source used, different under-stoichiometric \ce{Cs_{\textit{x}}Cr5Te8} phases with 0.73\,$<$\,\textit{x}\,$<$\,1 have been synthesized. Both structure types contain \ce{CrTe2} layers of edge-sharing \ce{CrTe6} octahedra, but the number of \ce{Cr2Te2} bridges connecting them controls the dimensionality -- from two-dimensional in delafossite-type to three-dimensional in hollandite-like phases. This increase in dimensionality is accompanied by a decreases with alkali content. A structure with alternating \ce{Cr2Te2} bridges and layers of alkali atoms, however, has not yet been reported and motivated our search for new two-dimensional magnetic materials.

In this work, we apply flux synthesis strategies to these alkali-chromium-tellurides, revealing a novel ladder-like hybrid framework with tunable dimensionality and magnetic ground states -- opening a new avenue in single-anion materials design, and bridging a logical gap in the (Rb,Cs)-Cr-Te phase space. Specifically, we obtained large crystals of the previously unreported layered phases \ce{Cs_{2.4}Cr8Te14} and \ce{Rb_{2.4}Cr8Te14}. These compounds exhibit the desired double-layered ladder-like architecture, which can be interpreted as a structural intergrowth phase consisting of the structural elements of both the hollandite- and the delafossite-like structures. In direction-dependent magnetization experiments on oriented single crystals we find that \ce{Rb_{2.4}Cr8Te14} is antiferromagnetic with $T_{N}$\,=\,114.5\,K, while \ce{Cs_{2.4}Cr8Te14} orders ferrimagnetically at \textit{T}\textsubscript{C} = 125\,K.

\section*{Experimental}
\label{experimental}

\subsection*{Synthesis of \ce{Cs_{2.4}Cr8Te14}}
As a first step, a Cs\textsubscript{\textit{x}}Te\textsubscript{\textit{y}} precursor was prepared by carefully adding drops of liquid Cs (Alfa Aesar, 99.98\%) to crushed Te pieces (Alfa Aesar, 99.999\%) in an agate mortar inside an Ar-filled glovebox, which led to the rapid formation of unspecified Cs--Te binary compounds, which were ground into a fine powder. Cr (powder, Alfa Aesar, 99.95\%) and the Cs\textsubscript{\textit{x}}Te\textsubscript{\textit{y}} precursor were then placed in a Canfield alumina crucible set comprising a bottom and top crucible and a frit disc in between \cite{Canfield2016}. The setup was then sealed in a quartz ampule under an atmosphere of 300\,mbar Ar. Inside a muffle furnace, the ampule was heated at 30\,°C/h to 1000\,°C, and was subsequently slow cooled to 750\,°C within 96\,h. After this, the ampule was removed from the oven, and the excess flux was separated by immediate high-temperature centrifugation.

\subsection*{Synthesis of \ce{Rb_{2.4}Cr8Te14}}
Single crystals of \ce{Rb_{2.4}Cr8Te14} were synthesized using a Rb/Se self flux starting from a molar Rb:Cr:Te ratio of 3.3:1:8. Rb (Strem, 99.9+\%), Cr (powder, Alfa Aesar, 99.95\%) and Te  (pieces, Sigma Aldrich, 99.999\%) were placed in a Canfield crucible inside a glovebox with Ar atmosphere. The crucible was then sealed in a quartz ampule under an atmosphere of 300\,mbar Ar and placed in a muffle furnace. The temperature program followed by the subsequent centrifugation was identical to the synthesis of \ce{Cs_{2.4}Cr8Te14}.

\subsection*{Synthesis of \ce{CsCr5Te8}}
\begin{figure}[h!]
    \centering
    \includegraphics[width=1\textwidth]{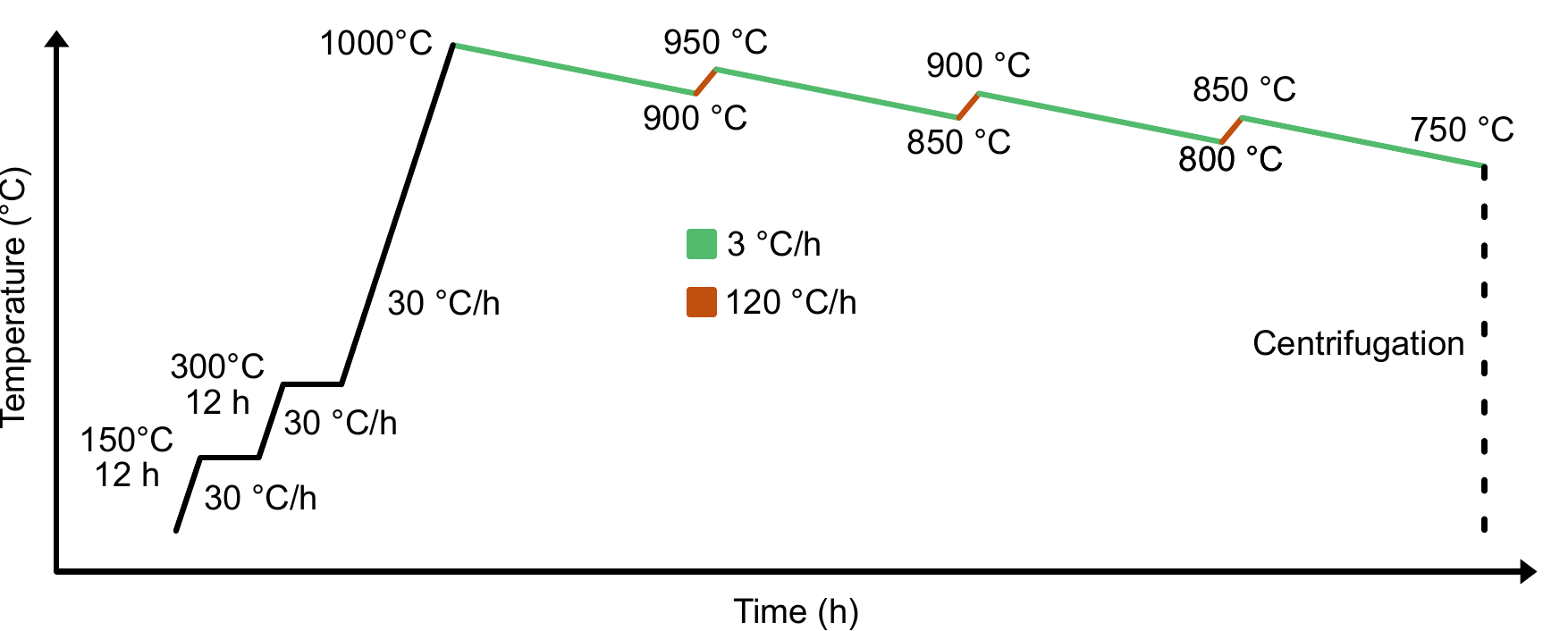}
    \caption{Adjusted temperature program used for the self-flux synthesis of larger single-crystals of \ce{CsCr5Te8} compared to two-step synthesis heating up to 1000\,°C with 30\,°C/h and subsequent cooling to 750\,°C over 96\,h. Axes not to scale.}
    \label{fig:TempProg}
\end{figure}

Single crystals of \ce{CsCr5Te8} were synthesized using a Cs/Te self-flux from molar Cs:Cr:Te ratios of 3.3:1:8 by the same procedure as \ce{Cs_{2.4}Cr8Te14}.
By this, black bar-shaped single crystals were obtained. Powder X-ray diffraction (PXRD) measurements confirmed an almost phase-pure product with small impurities of remaining \ce{Cs2Te3} and \ce{Cs2Te2} flux (see ESI).
By applying a saw-tooth like cooling profile (see, Figure \ref{fig:TempProg}), the crystal size of \ce{CsCr5Te8} platelets could be increased moderately.

\subsection*{Powder X-ray diffraction (PXRD)}
Capillary PXRD data of \ce{Rb_{2.4}Cr8Te14} and \ce{Cs_{2.4}Cr8Te14} were collected on a Bruker D8 Discover diffractometer equipped with a LynxeyeXE detector using Mo-K\textsubscript{$\alpha$} radiation ($\lambda$\,=\,0.710806\,\AA) in a 2$\theta$ scan in the range of 2.5--49°. The beam was shaped by a 6\,mm focussing Göbel mirror and size selected by a 0.6\,mm slit. Capillary PXRD measurements on \ce{CsCr5Te8} were performed on a Rigaku SmartLab diffractometer with Cu-K\textsubscript{$\alpha$} radiation ($\lambda$\,=\,1.54187\,\AA) on a D/teX Ultra 250 detector in the range of 5°--80°. For this, \ce{CsCr5Te8} was diluted with amorphous quartz dust. Calculated patterns were simulated using the Mercury software developed by the Cambridge Crystallographic Data Centre.\cite{Mercury} Full width half maximum (FWHM) values were estimated to match the experimental pattern best. Patterns of \ce{Rb2Te3}, \ce{Cs2Te3}, and \ce{Cs2Te2} were generated based on crystal structures from the ICSD database\cite{ICSD}: \ce{Rb2Te3} (90806), \ce{Cs2Te3} (53244), and \ce{Cs2Te2} (83351).

\subsection*{Single crystal X-ray diffraction (SXRD)} 
SXRD experiments were performed under \ce{N2} cooling at 100\,K on an Oxford Diffraction Supernova diffractometer using Mo K\textsubscript{$\alpha$} radiation ($\lambda$\,=\,0.71072\,\AA). Pre-experiment screenings, data collection, data reduction, and absorption correction were performed using the program suite CrysAlisPro.\cite{Rigaku2015} The crystal structure was solved with the dual space method in SHELXT.\cite{Sheldrick2015XT} Least squares refinements of F\textsuperscript{2} were performed using SHELXL\cite{Sheldrick2015XL}.

\subsection*{Magnetization measurements} Magnetization versus temperature and magnetization versus magnetic field measurements were carried out in a Physical Property Measurement System (PPMS DynaCool) from Quantum Design equipped with the vibrating sample magnetometer (VSM) option. The measurements were performed in a temperature range of 1.8 -- 300\,K in the sweep mode at rates of 5\,K$\cdot$min\textsuperscript{$-1$} and 50\,Oe$\cdot$s\textsuperscript{$-1$} in the range of -9\,T to 9\,T.\\

\subsection*{Scanning electron microscopy (SEM) and energy-dispersive X-ray spectroscopy (EDS)} Electron images were obtained from a JEOL JSM-IT800 Scanning electron microscope with an acceleration voltage of 20\,kV. Energy dispersive X-ray spectroscopy (EDS) data was collected with an X-Max\textsuperscript{N} 80 detector from Oxford Instruments. Composition measurements are based on 80 measurement points at eight sites on four crystals from two batches each.

\section*{Results and Discussion}
\label{results_discussion}

\subsection*{Crystal structures of \ce{Rb_{2.4}Cr8Te14} and \ce{Cs_{2.4}Cr8Te14}}

\begin{figure}[h!]
    \centering
    \includegraphics[width=1\textwidth]{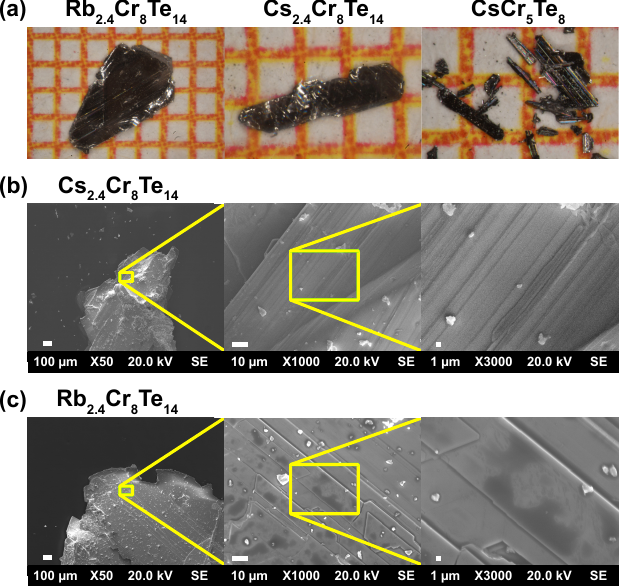}
    \caption{(a) Photographs of obtained crystals of \ce{Rb_{2.4}Cr8Te14}, \ce{Cs_{2.4}Cr8Te14}, and \ce{CsCr5Te8} taken on millimeter-sized graph paper. (b,c) SEM, SE (Secondary electron) images of (b) \ce{Cs_{2.4}Cr8Te14} and (c) \ce{Rb_{2.4}Cr8Te14}. Yellow frames indicate the frame of the respective image with increased magnification of $\times$50, $\times$1000, and $\times$3000 taken perpendicular to the crystal's surface.}
    \label{Photo_EDS}
\end{figure}

\begin{figure*}
\begin{center}
\includegraphics[width=17.4cm]{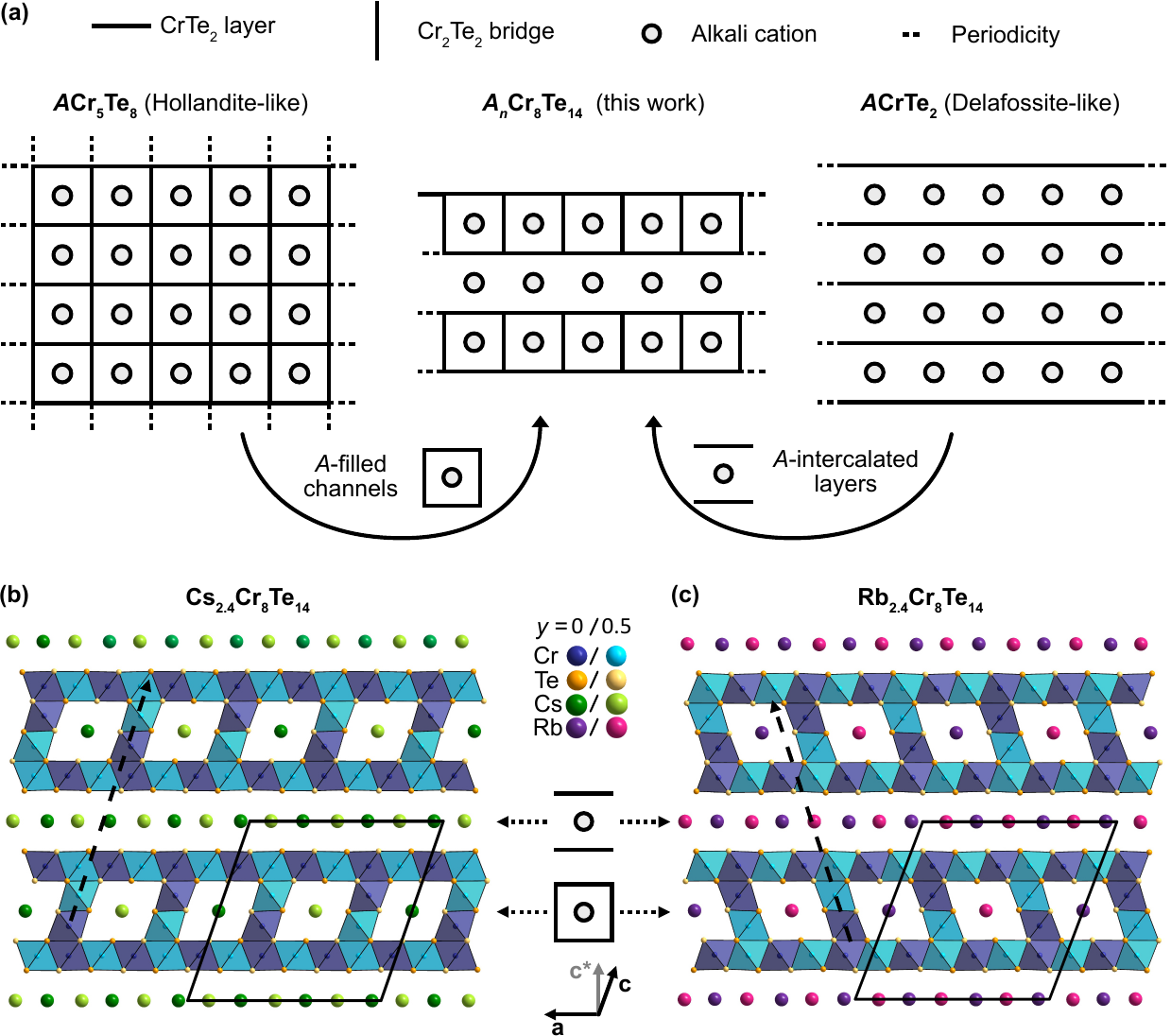}
\caption{(a) Schematic comparison of the crystal structures of hollandite-like \textit{A}\ce{Cr5Te8} and delafossite-like \textit{A}\ce{CrTe2} phases with \textit{A}\ce{_{2.4}Cr8Te14} compounds presented in this work. Arrows indicate the combination of the structural motives (filled tunnels and intercalated layers) found in \textit{A}\ce{_{2.4}Cr8Te14}. Crystal structures of (b) \ce{Cs_{2.4}Cr8Te14} and (c) \ce{Rb_{2.4}Cr8Te14} . Cr atoms are depicted in blue, Te in orange, Cs in green, and Rb in purple. Atoms located at y = 0 are depicted dark and atoms at y = 1/2 are depicted in light colors. Dashed arrows represent extensions of the \ce{Cr2Te2} bridges and underline the difference in stackings of layers.}
\label{fig:structure}
\end{center}
\end{figure*}

Metallic-black crystals of \ce{Rb_{2.4}Cr8Te14} (starting from a flux with molar Rb:Cr:Te ratios of 3.3:1:8) with sizes up to 4 $\times$ 3 $\times$ 0.2 mm$^{3}$ and \ce{Cs_{2.4}Cr8Te14} (starting from  a flux with molar Cs:Cr:Te ratios of 6:1:8; size up to 3 $\times$ 1 $\times$ 0.2 mm$^{3}$) were obtained from self-flux syntheses followed by hot-centrifugation (Photographs Figure \ref{Photo_EDS}). Powder X-ray diffraction (PXRD) identified \ce{Rb2Te3} or \ce{Cs2Te3} and \ce{Cs2Te2}, as the only detectable side products, resulting from residual flux material (see ESI). When the Cs-based synthesis is performed with a molar 3.3:1:8 ratio of the starting elements instead, crystals of the hollandite-like phase \ce{CsCr5Te8} are obtained; these were characterized by PXRD and thanks to increased crystal size subjected to anisotropic magnetic measurements for the first time. This demonstrates that in the (Rb,Cs)--Cr--Te system, the flux composition operates as a control parameter on phase formation. 

Our SXRD (Single-crystal X-ray diffraction) measurements of \ce{Rb_{2.4}Cr8Te14} and \ce{Cs_{2.4}Cr8Te14} revealed that both phases crystallize in new structure types. While they share the same space group $Cm$ and the same basic structural building blocks, they are nonetheless not isotypic due to a different stacking arrangement of the \ce{\textit{A}Cr8Te14} double layers. Figure \ref{fig:structure}(a) depicts the building blocks derived from the hollandite- and delafossite-like phases, which combine to form the ladder-like architecture characteristic of these compounds.

The lattice parameters of \ce{\textit{A}_{2.4}Cr8Te14} (\textit{A} = Rb, Cs) are very similar \textit{a} = 20.2039(5) \AA, \textit{b} = 3.91660(10) \AA, \textit{c} = 19.8178(6) \AA, $\beta$ = 109.152(3)°, \textit{V} = 1481.40(7) \AA$^{3}$ for \ce{Cs_{2.4}Cr8Te14} and \textit{a} = 20.1741(3) \AA, \textit{b} =  3.91560(10) \AA, \textit{c} = 19.6823(4) \AA, $\beta$ = 110.827(2)°, \textit{V} = 1453.19(6) \AA$^{3}$ for \ce{Rb_{2.4}Cr8Te14}. (See Table \ref{tbl:crydata}) The solution and refinement of both crystal structures was complicated by the soft mechanical behavior of the crystal, residual flux and several possible sources of disorder. These challenges are discussed in detail in the ESI.

In the crystal structures of \ce{\textit{A}_{2.4}Cr8Te14}, all atoms are located at crystallographic sites corresponding to the 2\textit{a} Wyckoff position with site symmetry \textit{m}, in total eight Cr, fourteen Te, and four \textit{A} sites. Like in other ternary \textit{A}--Cr--Te phases, \ce{CrTe2} layers, which consist of edge-sharing \ce{CrTe6} octahedra, form the structural backbone. 

In delafossite-like \ce{\textit{A}CrTe2} phases, the \ce{CrTe2} layers are intercalated by layers of \textit{A} cations (Figure \ref{fig:structure}(a)). In contrast, in \textit{A\textsubscript{x}}\ce{Cr5Te8} phases, these \ce{CrTe2} layers are instead interconnected by alternating \ce{Cr2Te2} bridges and \textit{A} cations. Through this, a tri-periodic Cr--Te framework permeated by channels along [010] is formed. In \ce{\textit{A}_{2.4}Cr8Te14}, both of these two possibilities of linking \ce{CrTe2} layers are realized alternately. Pairs of \ce{CrTe2} layers are interconnected by \ce{Cr2Te2} bridges leading to a ladder-shaped \ce{Cr8Te14} double-layer. Like their counterparts within the \ce{CrTe2} layers, the \ce{CrTe6} coordination polyhedra of the two bridging Cr atoms are connected to each other by edge-sharing. However, the contact between \ce{CrTe2} layers and \ce{Cr2Te2} bridges is facilitated by plane-sharing instead, thus resulting in much shorter Cr---Cr distances at these points. Instead of 3.8--4.0 \AA, the Cr---Cr contacts are as close as 3.078(15) and 3.087(15) \AA\, (Cs-phase) or 3.061(10) and 3.065(10) \AA\, (Rb-phase). 

The \ce{Cr2Te2} bridge further influence the Cr---Cr distances within the \ce{CrTe2} layers: contacts along $<$010$>$ are near 3.92 \AA, while those along $<$130$>$ alternate between shorter (3.8--3.9 \AA) and longer ($ \approx$ 4.0 \AA) separations. This distance disproportionation reflects the influence of the interconnecting \ce{Cr2Te2} bridges bound by plane-sharing, as the Cr sites with the most distorted in-plane Cr---Cr contacts are those directly involved in the cross-connections.

The \ce{Cr8Te14} ladders are isolated from each other by a layer of \textit{A} cations. While the \textit{A} sites located inside the hollandite-type tunnels are fully occupied, as they are in \ce{CsCr5Te8} as well, those forming the separating layer are occupationally disordered with site occupation factors close to 0.5. This leads to the non-integer sum formula of \ce{\textit{A}_{2.4}Cr8Te14}, as a full occupation of all \textit{A} sites would instead lead to a sum formula of \ce{\textit{A}_{4}Cr8Te14}. The \textit{A} content is very similar in the presented SXRD refinements of \ce{Rb_{2.4}Cr8Te14} and \ce{Cs_{2.4}Cr8Te14}. The \ce{\textit{A}_{2.4}Cr8Te14} stoichiometry of both phases was later confirmed by EDS measurements discussed in detail below.
Under-stoichiometry of the \textit{A} cations is frequently observed in delafossite-like phases, including the cathode materials \ce{Na_{\textit{x}}CoO2} and \ce{Li_{\textit{x}}CoO2}\cite{Hertz2008,Foo2004}, or the chromium based phases \ce{K_{0.6--0.8}CrSe2}\cite{Wiegers1980,Eder2025_2}, \ce{K_{0.87}CrSe2}\cite{Eder2025} and \ce{K_{1--\textit{x}}CrSe2}\cite{Song2021}, as well as in hollandite-like \ce{\textit{A}_{\textit{x}}Cr5Te8} (\textit{A} = Rb, Cs; 0.619 $< x <$ 1)\cite{Bouteiller2023}. In all cases, the removal of \textit{A} cations leads to a gradual increase of the formal Cr oxidation number from pure Cr\textsuperscript{III} to a mixed Cr\textsuperscript{III}/Cr\textsuperscript{IV} valence state. These interlayer \textit{A} sites are the only part of the crystal structure that breaks \textit{C}2/\textit{m} symmetry, which is followed by the \ce{Cr8Te14} ladders. The consequences of this on the refinement are discussed in detail in the ESI.

\subsection*{Differences between \ce{Cs_{2.4}Cr8Te14} and  \ce{Rb_{2.4}Cr8Te14}}

At first glance, the unit cell metrics and crystal structures of \ce{Cs_{2.4}Cr8Te14} and  \ce{Rb_{2.4}Cr8Te14} appear very similar. The \textit{a} and \textit{b} lattice parameters, which are mostly defined by the dimensions of the \ce{Cr8Te14} ladders, are almost identical. The larger value of \textit{c} for \ce{Cs_{2.4}Cr8Te14} can be attributed to the higher ionic radius of Cs compared to Rb. 

However, the most striking difference concerns the orientation of the \ce{Cr2Te2} bridges relative to \textbf{c}. While in \ce{Cs_{2.4}Cr8Te14}, they are located parallel to \textbf{c}, in \ce{Rb_{2.4}Cr8Te14}, they are tilted in the other direction relative to \textbf{c*}. This indicates that the stacking of the whole \ce{Cr8Te14} layers is different for the two phases (see Figure \ref{fig:structure}(b,c)). This difference cannot be resolved by a different setting of the unit cell. When progressing along the tilted direction of the \ce{Cr2Te2} bridges (dashed arrows in Figure \ref{fig:structure}(b,c)), the Cr atoms alternate pairwise between \textit{y} = 0 and \textit{y} = 1/2 in \ce{Cs_{2.4}Cr8Te14}, whereas in \ce{Rb_{2.4}Cr8Te14} this alternation follows a repeating sequence of \textit{y} = 0, 1/2, 1/2, 0 along the bridge direction. This clear difference in the stacking arrangement of the double-layers underlines the presence of two different structure types in \ce{Cs_{2.4}Cr8Te14}, and \ce{Rb_{2.4}Cr8Te14}.

A similar distinction between two closely related structure types is known for the pseudo-hollandite \ce{\textit{A}Cr5Te8} phases. There, the difference between the so-called A- and B-type structures\cite{Yamazaki2011} lies in whether neighboring \ce{Cr2Te2} bridges connect to the same \ce{CrTe6} octahedron (B) or from neighboring ones (A). 

\begin{table*}
	\begin{center}
	\caption{Crystallographic data for single-crystals of \ce{Cs_{2.4}Cr8Te14} and \ce{Rb_{2.4}Cr8Te14}.}
   \label{tbl:crydata}
		\begin{tabular}{lll}	
\toprule		
\multicolumn{3}{l}{\textbf{Physical, crystallographic, and analytical data}}\\
\midrule
& \textbf{\ce{Cs_{2.40}Cr8Te14}} & \textbf{\ce{Rb_{2.40}Cr8Tr14}}\\
Chemical formula & \ce{Cs_{2.40(2)}Cr8Te14}  & \ce{Rb_{2.40(2)}Cr8Tr14} \\
CCDC Deposition code & 2495117 & 2495116\\
Mol. wt. (g$\cdot$cm$^{-3}$) & 2520.91 & 2407.39\\ 
Cryst. syst. & monoclinic & monoclinic\\
Space group & \textit{Cm} (8) & \textit{Cm} (8) \\
$a$ (\AA) & 20.2039(5) & 20.1741(3)\\ 
$b$ (\AA) & 3.91660(10) & 3.91560(10)\\ 
$c$ (\AA) & 19.8178(6) & 19.6823(4)\\ 
$\beta$ (°) & 109.152(3) & 110.827(2)\\ 
$V$ (\AA$^{3}$) & 1481.40(7) & 1453.19(6)\\ 
$Z$ & 2 & 2\\
Calculated density (g$\cdot$cm$^{-3}$) & 5.652 & 5.502\\
Temperature (K) & 100(2) & 100(2)\\
Diffractometer & \multicolumn{2}{l}{Rigaku Oxford Diffraction SuperNova} \\
Radiation ($\lambda$) & \multicolumn{2}{l}{Mo K$\alpha$ (0.71073 \AA)} \\
Crystal color & metallic black & metallic black\\
Crystal description & plate & plate\\
Crystal size (mm$^{3}$) & 0.16 $\times$ 0.12 $\times$ 0.02 & 0.19 $\times$ 0.12 $\times$ 0.02 \\
Linear absorption coefficient (mm$^{-1}$) & 19.225 & 20.561\\
Scan mode & \multicolumn{2}{l}{$\omega$} \\
Recording range $\theta$ (°) & 2.499 to 30.570 & 2.484 to 33.0737\\
$h$ range & --28 $\le h \le$ +28 & --29 $\le h \le$ 30\\
$k$ range & --5 $\le k \le$ +5 & --5 $\le k \le$ 5\\
$l$ range & --28 $\le l \le$ +28 & --29 $\le l \le$ 29\\
Nr. of measured reflections & 44640 & 47500\\
[1ex]\\
\multicolumn{3}{l}{\textbf{Data reduction}} \\
Completeness (\%) & 99.7 & 99.4 \\
Nr. of independent reflections & 5144 & 5796\\
$R_{\mathrm{int}}$ (\%) & 5.39 & 4.08 \\
$R_{\mathrm{\sigma}}$ (\%) & 3.14 & 2.37 \\
Absorption correction & \multicolumn{2}{l}{numerical (Gaussian grid)}\\
Independent reflections \\ with I $\geq$ 2.0$\sigma$ & 3601 & 4595\\
[1ex]\\
\multicolumn{3}{l}{\textbf{Refinement}}\\
$R1$ (obs / all) (\%) & 6.43 / 8.90 & 4.27 / 5.64 \\
$wR2$ (obs / all) (\%) & 17.92 / 19.83 & 12.12 / 13.40 \\
$GOF$ & 1.106 & 1.074\\
Nr. of refined parameters & 117 & 117\\
Nr. of restraints & 2 & 2\\
Difference Fourier residues ($e^-$\AA$^{-3}$) & --3.64, 7.53 & --2.63, 5.43 \\
\bottomrule	
	\end{tabular}
	\end{center}

\end{table*}

\subsection*{Microstructure and elemental composition}
EDS analysis was performed on \ce{Cs_{2.4}Cr8Te14} and \ce{Rb_{2.4}Cr8Te14} crystals to verify the elemental composition, especially the alkali content determined by the SXRD experiments. EDS measurements based on 80 measurement points at eight sites on four crystals from two batches of search found the elemental composition to be Cs\textsubscript{2.9(4)}Cr\textsubscript{8}Te\textsubscript{14.6(8)}, which shows that the Cs and Te contents are in very good agreement with our SXRD experiment. The elementary composition of \ce{Rb_{2.4}Cr8Te14} based on EDS was found to be  Rb\textsubscript{2.56(20)}Cr\textsubscript{8}Te\textsubscript{13.7(3)} which fits the values from SXRD within one standard deviation. Both the alkali metal, as well as the Te contents are slightly elevated in comparison with the refined structural model from SXRD, which might be caused by some residual flux stuck to the uncleaved sample surface. According to EDS analyses, \ce{Cs_{2.4}Cr8Te14} appears to contain more excess flux, which is in line with traces of \ce{Cs2Te2} observed in PXRD, which is expected to be solid at the centrifugation temperature.\cite{Okamoto_2016_Cs2Te2} It should be noted that the chemical composition, especially for \ce{Rb_{2.4}Cr8Te14}, was found not to vary much when comparing different spots on the same crystal, different crystals as well as crystals from different batches. This suggests homogeneity of the composition within the crystals and the batches as well as the reproducibility of the synthesis. Maps of the elemental compositions with two different magnifications further underlining the homogeneity of the crystals are depicted in the ESI.The chemical composition of flux-grown \ce{CsCr5Te8} was determined as Cs\textsubscript{1.01(11)} Cr\textsubscript{5}Te\textsubscript{8.1(4)}, which exactly fits the expected stoichiometry and does not indicate any under-occupation of the Cs sites in the channels. A list of the individual elemental compositions within the batches, crystals, and sites for all three compounds is given in the ESI.

Using scanning electron microscopy, we further studied the microstructure of \ce{Cs_{2.4}Cr8Te14} and \ce{Rb_{2.4}Cr8Te14}, the results of which are depicted in Figure \ref{Photo_EDS}(b,c). Both compounds form overall plate-shaped crystals that show large areas of flat surfaces. Large magnifications show steps suggesting stacks of multilayered, stacked crystal slabs, which is a common microstructural feature of 2D-materials. On top of the surface, smaller on-grown crystals can be observed, which were identified as remaining flux $A_{x}$Te$_{y}$ with no Cr-content. The layered nature of the crystal structure is further highlighted by stair-like features resulting from the increasing number of multi-layers on top of each other. 

\subsection*{Magnetic properties}

\begin{figure*}
\begin{center}
\includegraphics[width=17.4cm]{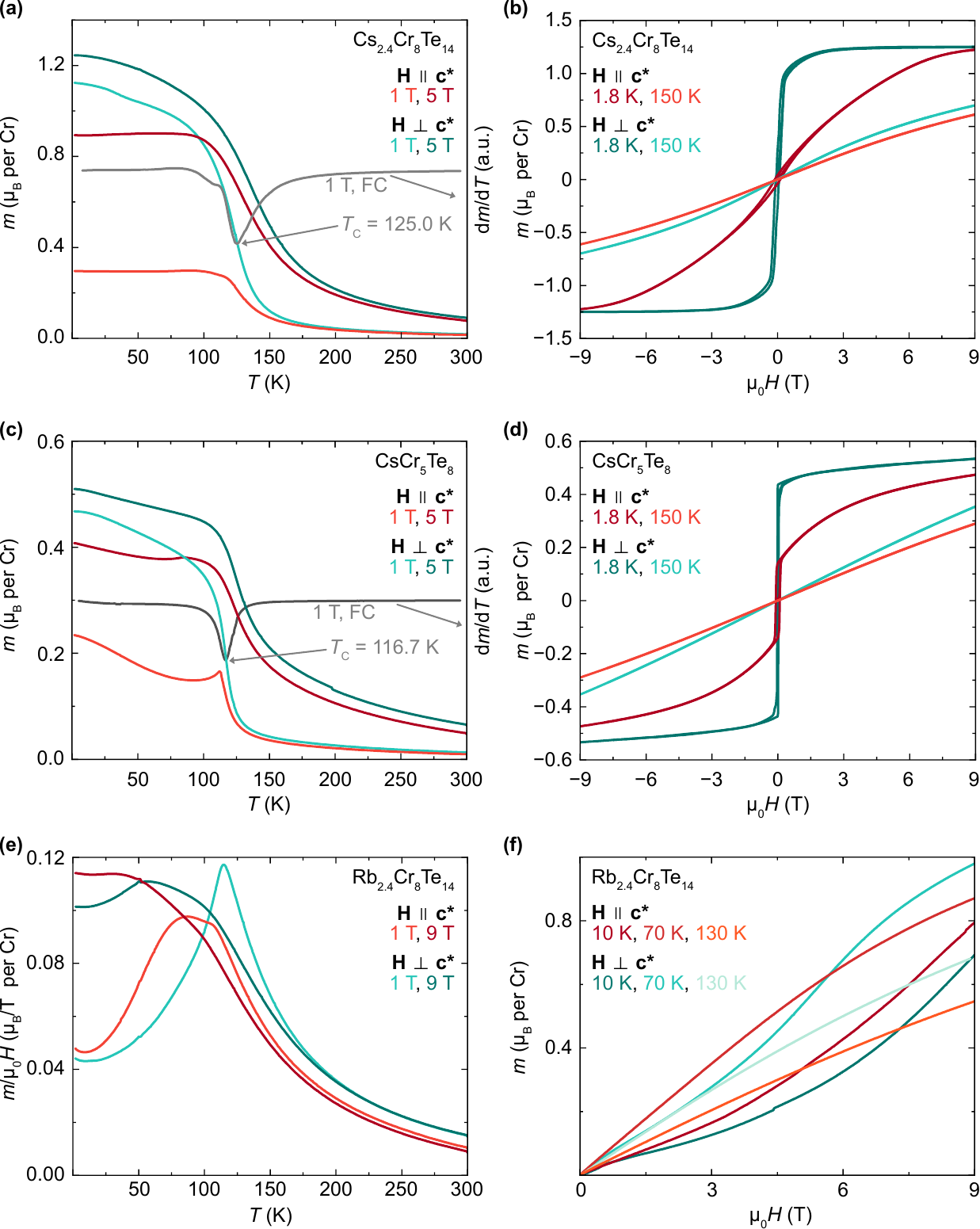}
\caption{Temperature-dependent and field-dependent magnetic moment of \ce{Cs_{2.4}Cr8Te14}, \ce{CsCr5Te8}, and \ce{Rb_{2.4}Cr8Te14}. Field orientations are indicated by color: Parallel to \textbf{c*} axis are in red and perpendicular to it in cyan. 
(a,c,e) Field-cooled data between 1.8\,K and 300\,K. (a) \ce{Cs_{2.4}Cr8Te14} at 1\,T (+ derivative) and 5\,T. (c) \ce{CsCr5Te8} at 1\,T (+ derivative) and 5\,T. (e) \ce{Rb_{2.4}Cr8Te14} at 1\,T and 9\,T.
(b,d,f) Field-dependent measurements: (b) \ce{Cs_{2.4}Cr8Te14} at 1.8\,K and 150\,K between --9\,T and 9\,T. (d) \ce{CsCr5Te8} at 1.8\,K and 150\,K between --9\,T and 9\,T. (f) \ce{Rb_{2.4}Cr8Te14} at 10\,K, 70\,K and 130\,K up to 9\,T. Derivatives are based on the 1\,T FC measurement in the \textbf{H} $\perp$ \textbf{c*} orientation.}
\label{fig:Magnetism}
\end{center}
\end{figure*} 

Given the uncommon isolated ladder-like architecture of these compounds, their magnetic behavior is of particular interest, and we have studied it using direction-dependent magnetization on oriented single crystals. This is especially relevant in light of the rich and tunable magnetism known in structurally related intercalated chromium dichalcogenides, where small variations in size and amount of alkali cations lead to diverse magnetic ground states. Analogously, the closely related ladder-like telluride frameworks of \ce{Rb_{2.4}Cr8Te14} and \ce{Cs_{2.4}Cr8Te14} provide an excellent platform to explore the interplay between crystal structure and magnetism. As shown in Figure \ref{fig:Magnetism}, the magnetic moment reveals that the two compounds, despite their structural similarity, order magnetically in different ways: \ce{Rb_{2.4}Cr8Te14} orders antiferromagnetically at $T_{\mathrm{N}}$ = 114.5 K, whereas \ce{Cs_{2.4}Cr8Te14} exhibits ferrimagnetic ordering at $T_{\mathrm{C}}$ = 125.0 K.\\
Temperature-dependent measurements of the magnetic moment of \ce{Cs_{2.4}Cr8Te14} (Figure \ref{fig:Magnetism} (a)) show a transition to a ferromagnetic state below $T_{\mathrm{C}}$ = 125.0\,K. The ferromagnetic transition temperature can be estimated based on the maxima of the first derivative $\frac{dm}{dT}$ corresponding to the inflection point in the \textit{m}(\textit{T}) curves at $T_{\mathrm{C}}$ = 125.0\,K. 

The field-dependent magnetization measurements of \ce{Cs_{2.4}Cr8Te14} are presented in Figure \ref{fig:Magnetism}(b) for $T$ = 1.8\,K, and 150\,K, with the external magnetic field parallel and perpendicular to \textbf{c*} respectively. Only a small hysteresis loop is visible at T = 1.8 K. For both crystal orientations in the external field, \textbf{H}\,$\parallel$\,\textbf{c*} = 9\,T and \textbf{H} $\perp$ \textbf{c*} = 9\,T the magnetic moment at 1.8\,K saturates at about $m$ = 1.24\, $\mu_{B}$/Cr. Since the magnetic moment saturates much faster, when the external field is applied parallel to the layer plane (\textbf{H}\,$\perp$\,\textbf{c*}), the magnetic easy-axis and direction of the magnetic moments are likely to lie within the layer plane as well. The theoretical magnetic moment can be estimated using the spin-only formula $m = g\,\cdot\,S\,\cdot\,$\textmu\textsubscript{B} (assuming no spin-orbit coupling) with $S$ being the total spin and $g$ being the Lande factor, which can be approximated as being 2. Since the formal oxidation number of Cr in \ce{\textit{A}_{2.4}Cr8Te14} amounts to +3.2, we estimate the magnetic moment from a mixture of the adjacent +3 and +4 oxidation states. For Cr\textsuperscript{4+} and Cr\textsuperscript{3+} the expected moments are 2\,\textmu\textsubscript{B} and 3\,\textmu\textsubscript{B} respectively, leading to a theoretically expected moment of 2.8 \textmu\textsubscript{B}/Cr atom, which is significantly higher than the observed saturation moment. This difference can be rationalized by ferrimagnetic ordering within the \ce{Cr8Te14} ladders, instead of a pure ferromagnetic ground state. The two types of \ce{CrTe6} octahedra-linking by edge- and face-sharing and the resulting two groups of nearest-neighbor Cr---Cr-distances provide the structural basis for such a magnetic ordering. According to the Goodenough--Kanamori--Anderson rules,\cite{Anderson1950_GKA,Goodenough1955_GKA,Goodenough1958_GKA,Kanamori1957_GKA_1,Kanamori1957_GKA_2} the 
edge--sharing \ce{CrTe6} octahedra with Cr--Te--Cr angles close to 90$^\circ$ favor ferromagnetic superexchange, while the Cr\textsuperscript{III}---Cr\textsuperscript{III} contacts evoke antiferromagnetic direct exchange; the latter contribution becomes increasingly relevant as the Cr---Cr distance decreases.
Within the \ce{CrTe2} layers, the large Cr---Cr distances of 3.8--4.0 \AA\,make ferromagnetic interactions favorable. In contrast, the short Cr---Cr contacts (< 3.1 \AA) that connect the \ce{Cr2Te2} slabs to the \ce{CrTe2} layers are very likely antiferromagnetically coupled. Assuming two spin orientations along the easy axis, this geometry would couple the two \ce{CrTe2} layers ferromagnetically while the bridging \ce{Cr2Te2} units adopt the opposite spin orientation. With six Cr atoms per formula unit in the \ce{CrTe2} layers and two in the bridges, the net moment is reduced by a factor of $\frac{6-2}{8}=\frac{1}{2}$, which is consistent with the reduced saturation magnetization observed in \ce{Cs_{2.4}Cr8Te14}.\\
This hypothesis is underpinned by the magnetic saturation moments of \ce{CsCr5Te8} observed in the magnetic field-dependent measurements in Figure \ref{fig:Magnetism}(d). The saturation moment for an applied magnetic field \textbf{H} $\perp$ \textbf{c*} (along the easy magnetization axis) corresponds to slightly above 0.5\,\textmu$_{B}$. Due to the same distribution of Cr---Cr distances, we can rationalize the reduced saturation value again with ferromagnetically coupled \ce{CrTe2} layers and \ce{Cr2Te2} bridges exhibitng the opposite spin orientation. 
With three Cr atoms per formula unit in the layers and two in the bridges, the net moment is reduced to a factor of $\frac{3-2}{5}=\frac{1}{5}$. The expected value of 0.56\,\textmu$_{B}$/Cr is in excellent agreement with our observed $\sim$0.5\,\textmu$_{B}$/Cr. The Curie temperature of our flux-grown \ce{CsCr5Te8} was determined at $T$\textsubscript{C} = 116.7\,K based on the temperature-dependent magnetic moment data displayed in Figure \ref{fig:Magnetism}(c), which is slightly lower than previously reported (125\,K).\cite{Yamazaki2011} A more detailed magnetic investigation of \ce{CsCr5Te8} can be found in the ESI.

Temperature-dependent measurements of the magnetic moment normalized to the applied field of \ce{Rb_{2.4}Cr8Te14} are illustrated in Figure \ref{fig:Magnetism}(e). For an applied field of 1\,T, the magnetic moment shows the characteristic peak shape of an antiferromagnet with a maximum reached at the Néel temperature $T$\textsubscript{N} = 114.5\,K. This maximum is sharp for $\mu_0H$ = 1\,T and \textbf{H} $\perp$ \textbf{c*}, but becomes broadened for \textbf{H} $\parallel$ \textbf{c*} and increasing field strengths. The plateauing of \textit{m}(\textit{T}) at about 1\,$\mu_{B}$/Cr at $\mu_0H$ = 9\,T with \textbf{H} $\parallel$ \textbf{c*} indicates a shift to a spin-polarized ferromagnetic state. Figure \ref{fig:Magnetism}(d) shows the magnetic moment for applied fields between 0\,T and 9\,T at temperatures of 10\,K, 70\,K and 130\,K with fields applied \textbf{H} $\perp$ \textbf{c*} and \textbf{H} $\parallel$ \textbf{c*} respectively. Other than for \ce{Cs_{2.4}Cr8Te14}, no saturation was observed up to fields of 9 T in either crystal orientation. 
With increasing temperature, the curvature of \textit{m}(\textit{H}) changes from positive to negative, indicating a signature of a metamagnetic transition like a spin-flop. The change of curvature appears at different temperatures, approximately around 100 K for \textbf{H} $\perp$ \textbf{c*}, while already at ca. 60 K for \textbf{H} $\parallel$ \textbf{c*}. Supplementary magnetic analyses of all three phases are collated in the ESI.

\section*{Conclusion}
\label{conclusion}
	
We have extended the family of alkali chromium tellurides by discovering two phases with new structure types, \textit{A}\textsubscript{2.4}\ce{Cr8Te14} (\textit{A} = Rb, Cs). These compounds were obtained as millimeter-sized single crystals via \textit{A}/Te self-flux syntheses, highlighting the potential of flux growth not only for producing high--quality crystals but also as a materials discovery tool. 

Their crystal structures consist of a ladder-like hybrid framework that integrates motifs from both delafossite-like \textit{A}\ce{CrTe2} and hollandite-like \textit{A}\ce{Cr5Te8} phases, representing a flux-guided intergrowth of 2D and 3D endmembers, and thus highlighting the possibility of a new structure series from these building blocks.

Our large single crystals enabled us to perform anisotropic magnetic measurements of these phases. Despite their close structural relationship, the two compounds exhibit distinctively different magnetic ground states: \ce{Rb_{2.4}Cr8Te14} undergoes antiferromagnetic ordering below $T_{\mathrm{N}}$ = 114.5 K with field-induced metamagnetic transition at higher fields, while \ce{Cs_{2.4}Cr8Te14} exhibits a ferrimagnetic ground state with $T_{\mathrm{C}}$ = 125.0 K and reduced saturation magnetization. These findings underscore the sensitivity of magnetic order to subtle structural and electronic differences within the ladder framework. The competition between ferromagnetic and antiferromagnetic interactions at different Cr---Cr distances drives the observed long-range magnetic orders.

By extending intergrowth concepts to magnetic chromium-chalchogenides through simple flux synthesis control, this study opens new directions for targeted synthesis in magnetic 2D quantum materials research. The layered architecture of \textit{A}\textsubscript{2.4}\ce{Cr8Te14} provides a promising foundation for further structural series of intergrown alkali-chalchogenide ternaries, as well as for manipulation by deintercalation, metathesis, or intercalation chemistry, offering a pathway to tune their dimensionality and electronic states. Such chemically engineered derivatives may unlock access to new families of two-dimensional spintronic or quantum materials.

\section*{Acknowledgements}

This work was supported by the Swiss National Science Foundation under grants No. PCEFP2\_194183 and No. 200021-204065. The authors would like to thank Céline Besnard and Pascal Schouwink for their help on capillary PXRD measurements.

\section*{Conflict of Interest}

There are no conflicts to declare.

%%%%%%%%%%%%%%%%%%%%%%%%%%%%%%%%%%%%%%%%%%%%%%%%%%%%%%%%%%
%%%%%%%%%%%%%%%%%%%%%%%%%%%%%%%%%%%%%%%%%%%%%%%%%%%%%%%%%%
%%%%%%%%%%%%%%%%%%%%%%%%%%%%%%%%%%%%%%%%%%%%%%%%%%%%%%%%%%
\begin{shaded}
\noindent\textsf{\textbf{Keywords:} \keywords} 
\end{shaded}
%%%%%%%%%%%%%%%%%%%%%%%%%%%%%%%%%%%%%%%%%%%%%%%%%%%%%%%%%%
%%%%%%%%%%%%%%%%%%%%%%%%%%%%%%%%%%%%%%%%%%%%%%%%%%%%%%%%%%
%%%%%%%%%%%%%%%%%%%%%%%%%%%%%%%%%%%%%%%%%%%%%%%%%%%%%%%%%%

%%%%%%%		References			%%%%%%% 

\setlength{\bibsep}{0.0cm}
\bibliographystyle{Wiley-chemistry}
\bibliography{A2dot4Cr8Te14}

\clearpage

%%%%%%%		TOC Entry			%%%%%%% 

\section*{Entry for the Table of Contents}

%	 please select one option only and delete the other one

%%%%%%%		Option 1			%%%%%%%    
\noindent\rule{11cm}{2pt}
\begin{minipage}{5.5cm}
\includegraphics[width=5.5cm]{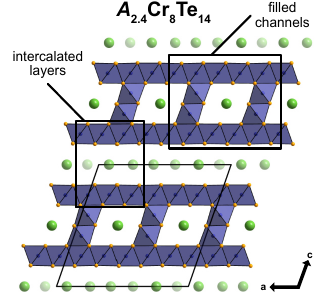} 
\end{minipage}
%\hspace{0.5cm}
\begin{minipage}{5.5cm}
\vspace{0.5cm}
\large\textsf{\textit{A}\ce{_{2.4}Cr8Te14} (\textit{A} = Rb, Cs) compounds exhibit a layered, ladder-like Cr-Te framework uniting structural motives of hollandite-like and delafossite-like phases. Alkali atoms are situated both in the interlayer space and tunnels within the \ce{Cr8Te14} framework. Despite similar structures, different alkali atoms give rise to distinct magnetic ground states.}
\end{minipage}
\noindent\rule{11cm}{2pt}

%\noindent\rule{11cm}{2pt}
%\begin{minipage}{5.5cm}
%\includegraphics[width=5.5cm]{TOC_opt1.png} 
%\end{minipage}
%\hspace{0.5cm}
%\begin{minipage}{5.5cm}
%\large\textsf{Authors should provide a short Table of Contents graphical abstract and accompanying text (up to 450 characters including spaces). The graphical abstract should stimulate curiosity. Repetition or paraphrasing of the title and experimental details should be avoided.}
%\end{minipage}
%\noindent\rule{11cm}{2pt}

%\vspace{2cm}

%%%%%%%		 Option 2			%%%%%%%    

%\noindent\rule{11cm}{2pt}
%\begin{minipage}{11cm}
%\includegraphics[width=11cm]{TOC_opt2.png}
%\end{minipage}
%\begin{minipage}{11cm}
%\large\textsf{Authors should provide a short Table of Contents graphical abstract and accompanying text (up to 450 characters including spaces). The graphical abstract should stimulate curiosity. Repetition or paraphrasing of the title and experimental details should be avoided.}
%\end{minipage}
%\noindent\rule{11cm}{2pt}

%%%%%%%%%%%%%%%%%%%%%%%%%%%%%%%%%%%%%%%%%%%%%%%%%%%%%%%%%%
%%%%%%%%%%%%%%%%%%%%%%%%%%%%%%%%%%%%%%%%%%%%%%%%%%%%%%%%%%
%%%%%%%%%%%%%%%%%%%%%%%%%%%%%%%%%%%%%%%%%%%%%%%%%%%%%%%%%%

\end{document}